\documentclass[review]{elsarticle}
\usepackage{amsmath}
\usepackage{graphicx}
\usepackage{epsfig}
\usepackage{psfrag}
\usepackage{array}
\usepackage{amssymb}
\usepackage{color}

\usepackage{amsmath,epsfig,bbm,amsbsy,natbib,amssymb,amsthm}
\usepackage{subfigure,psfrag,graphicx,amssymb,color}
\usepackage{graphicx,amssymb}
\usepackage[normalem]{ulem}
\usepackage{multirow, rotating}

\usepackage{stfloats}

\begin{document}

\begin{frontmatter}

\title{Reweighted $l_1$-norm Penalized LMS for Sparse Channel
Estimation and Its Analysis}
\author[uofa]{Omid Taheri}
%\ead{otaheri@ualberta.ca}
\author[uofa,aalto]{Sergiy A.~Vorobyov}
%\ead{svor@ieee.org}
\address[uofa]{Dept. Elec. and Comp. Eng., 
University of Alberta,
 Edmonton, AB T6G 2V4 Canada}
\address[aalto]{School of Electrical Engineering, Aalto University, 
FI-00076 AALTO, Finland}

%\vspace{-.5cm}
\begin{abstract}
A new reweighted $l_1$-norm penalized least mean square (LMS) algorithm 
for sparse channel estimation is proposed and studied in this paper. 
Since standard LMS algorithm does not take into account the sparsity 
information about the channel impulse response (CIR), sparsity-aware 
modifications of the LMS algorithm aim at outperforming the standard 
LMS by introducing a penalty term to the standard LMS cost function 
which forces the solution to be sparse. Our reweighted $l_1$-norm 
penalized LMS algorithm introduces in addition a reweighting of the 
CIR coefficient estimates to promote a sparse solution even more and 
approximate $l_0$-pseudo-norm closer. We provide in depth quantitative 
analysis of the reweighted $l_1$-norm penalized LMS algorithm. An 
expression for the excess mean square error (MSE) of 
the algorithm is also derived which suggests that under the right 
conditions, the reweighted $l_1$-norm penalized LMS algorithm outperforms 
the standard LMS, which is expected. However, our quantitative analysis 
also answers the question of what is the maximum sparsity level in the 
channel for which the reweighted $l_1$-norm penalized LMS algorithm is 
better than the standard LMS. Simulation results showing the better 
performance of the reweighted $l_1$-norm penalized LMS algorithm 
compared to other existing LMS-type algorithms are given.
\end{abstract}

\begin{keyword}
Channel estimation \sep Gradient descent \sep Least mean square (LMS) 
\sep Sparsity.
\end{keyword}
\end{frontmatter}

\section{Introduction}
The least mean square (LMS) algorithm is very well known in the
field of adaptive signal processing \cite{WidrowStearns85},
\cite{Haykin02}. It belongs to the class of stochastic gradient
algorithms. The attractive feature of the LMS algorithm is that it 
does not need extensive stochastic knowledge of the channel and the
input data sequence unlike some other parameter estimation
methods such as the recursive least squares (RLS) and Kalman
filter. While RLS and Kalman filter need to know the covariance 
matrix of the input data sequence, the LMS algorithm only requires 
an approximate estimate of the largest eigenvalue of the covariance 
matrix for proper selection of the step size that guarantees the
convergence. The LMS algorithm is being employed in a wide variety 
of applications in signal processing and communications including
system identification \cite{Bershadetal08}, echo cancellation
\cite{Raoetal09}, channel estimation \cite{Colerietal02}, adaptive
communication line enhancement \cite{Macchietal91}, etc. A
particular application considered in this paper is that of
estimating a finite impulse response (FIR) channel. The choice of
the channel estimation algorithm for use in a communication system
comes down to the available information about the statistics of
the system, the desired performance of the estimation algorithm,
as well as the complexity of the estimation process.

The standard recursive parameter estimation algorithms do not
assume any information about the specific structure of the channel
being estimated. However, being aware of the channel structure one
can modify the standard algorithms in order to have a better
estimate of the channel. In this paper, we are concerned with a
class of channels where the channel impulse response (CIR) is
sparse. A time sparse discrete-time signal is the one with only a
few nonzero entries. In general, the domain that the signal is
sparse in does not necessarily have to be the time domain. Other
sparsity bases can also be used and are represented by an $N
\times N$ orthogonal matrix where $N$ is the length of the signal.

Sparsity-aware modifications of the LMS algorithm have been 
presented in the signal processing literature in the past few years. 
The methods introduced in \cite{Chenetal09}, \cite{Guetal09} add a 
penalty term to the standard LMS error function which is designed in 
a way to force the solution to be sparse. A penalty in the form of 
the $l_0$-pseudo-norm of the CIR is used in \cite{Guetal09}, while 
\cite{Chenetal09} uses the $l_1$-norm. In \cite{ShiShi10}, the mean 
square convergence and stability analysis for one of the algorithms 
in \cite{Chenetal09} for the case of white input signals is presented. 
A performance analysis of the $l_0$-pseudo-norm constraint LMS 
algorithm of \cite{Guetal09} is given in \cite{Suetal12}. In 
\cite{ShiMa10}, \cite{YangSobelman10}, variations of the algorithms 
in \cite{Chenetal09} are introduced. In \cite{ShiMa10}, the filter 
coefficients are updated in a transform domain which leads to faster 
convergence for non white inputs. In \cite{Murakamietal10}, the idea 
of using a weighted $l_1$-norm penalty for the purpose of sparse 
system identification is presented without any convergence analysis.
Moreover, sparsity promoting partial update LMS algorithms have been 
recently developed in \cite{PartUpdate}.

The authors of \cite{BershadBist05} introduce a scheme
that employs two sequential adaptive filters for communication
line or network echo cancelers. The method exploits the
sparseness of the CIR and uses two sequential LMS type structures
which are both shorter than the largest delay of the channel. A
family of the so called natural gradient estimation algorithms is
also studied in \cite{Martinetal02}. It is shown that the class of
sparse LMS algorithms presented has faster convergence rate.

Sparse diffusion schemes are presented in \cite{Chouvardasetal12} 
and \cite{DilorenzoSayed13} that provide adaptive algorithms for 
distributed learning in networks. In \cite{Chouvardasetal12}, 
projection methods over hyperslabs and weighted $l_1$-balls are 
presented and analyzed for distributed learning. Penalized cost 
functions are used in \cite{DilorenzoSayed13} to enforce the 
sparsity of the solution. Among the penalty terms considered is 
the weighted $l_1$-norm penalty of \cite{Chenetal09}. Convergence 
analysis for the distributed adaptive algorithm is also given in 
\cite{DilorenzoSayed13} for a convex penalty term.

Other channel estimation algorithms have also been modified to
either better adapt to a sparse channel or achieve the same
performance as the corresponding standard algorithms with lower
complexity. Time and norm-weighted least absolute shrinkage and
selection operator (LASSO) where weights obtained from RLS
algorithm has been presented in \cite{Angelosanteetal10}. A greedy
RLS algorithm designed for finding sparse solutions to linear
systems has been presented in \cite{Dumitrescuetal12}, and it has
been demonstrated that it has better performance than the standard
RLS algorithm for estimating sparse time-varying FIR channels. A
compressed sensing (CS)-based Kalman filter has been developed in
\cite{Vaswani08} for estimating signals with time varying sparsity
pattern.\footnote{CS is the theory that considers the problem of
sparse signal recovery from a few measurements
\cite{CandesWakin08}, \cite{Donoho06b}. The number of measurements
in CS is a lot smaller than the overall dimension of the signal.}

In this paper, we first derive the reweighted $l_1$-norm penalized
LMS algorithm which is based on modifying the LMS error
(objective) function by adding the $l_1$-norm penalty term and
also introducing a reweighting of the CIR
coefficients.\footnote{Some preliminary results (the method and
some simulation results) have been reported in the conference
contribution \cite{OmidSergiy11}.} Then the main contribution
follows that is the in depth study of the convergence and excess
mean square error (MSE) analysis of the reweighted $l_1$-norm
penalized LMS algorithm. It is worth mentioning that the analytic 
arguments in \cite{DilorenzoSayed13} can be applied to a 
centralized learning problem as well as a diffusion network. 
In this way, it is also possible to prove the mean square stability 
of the reweighted $l_1$-norm penalized LMS algorithm in a different 
manner than presented in this paper. Our simulation results show that 
the proposed algorithm outperforms the standard LMS as well as the
penalized sparsity-aware LMS algorithms of \cite{Chenetal09} and
approve our theoretical studies.

The rest of the paper is organized as follows. Section~\ref{sec:prelim} 
reviews the system model used and the standard LMS algorithm. In
Section~\ref{sec:rl1penallms}, the reweighted $l_1$-norm penalized
LMS algorithm is introduced. An analytical study of the
convergence of the reweighted $l_1$-norm penalized LMS algorithm
as well as its excess MSE is given in
Section~\ref{ssec:reweightedl1penalizedlmsconvergence}. Simulation
results comparing the performance of different sparsity-aware LMS
algorithms are given in Section~\ref{sec:SimRes}.
Section~\ref{sec:Conc} concludes the paper.

%\newpage
\section{System Model and Preliminaries}
\label{sec:prelim}

\subsection{Standard LMS}
\label{ssec:standrdlms}
The standard LMS algorithm is used to estimate the actual CIR 
of a system where the CIR vector denoted as $\boldsymbol{w}$. 
Let us introduce as well other notations that we need in the 
following. An estimate of the actual CIR vector $\boldsymbol{w}$ 
at the time step $k$ is denoted as $\boldsymbol{w}_k$. The 
system's input data vector is $\boldsymbol{x}_k$, $n_k$ stands 
for the additive noise, $d_k$ is the desired response of a 
system, and $e_k$ is the error signal. The CIR is
assumed to be of length $N$, and therefore, $\boldsymbol{w}
\triangleq [w_1 \; w_2 \; \cdots \;w_N]^T$, $\boldsymbol{w}_k
\triangleq [w_{1,k} \; w_{2,k} \; \cdots \;w_{N,k}]^T$, and
$\boldsymbol{x}_k \triangleq [x_k \; x_{k-1} \; \cdots
\;x_{k-N+1}]^T$, where $(\cdot)^T$ stands for the vector
transposition. As shown in Fig.~\ref{fig:comsystemmodel}
\begin{align}
d_k = \boldsymbol{w}^T \boldsymbol{x}_k + n_k \nonumber \\
e_k \triangleq d_k - \boldsymbol{w}^T_k \boldsymbol{x}_k .
\label{eq:comsysequations}
\end{align}
The noise samples $n_k$ are assumed to be independent and
identically distributed (i.i.d.) with zero mean and variance of
$\sigma^2_{\rm n}$. Also, the input data sequence
$\boldsymbol{x}_k$ and the additive noise samples $n_k$ are
assumed to be independent.

In standard LMS, the cost function is $L_k \triangleq (1/2)
e^{2}_k$, and it is minimized using the gradient descent algorithm
\cite{WidrowStearns85}. The update equation of the standard LMS
algorithm can be derived from the above mentioned cost function as
\begin{equation}
\boldsymbol{w}_{k+1} = \boldsymbol{w}_k - \mu \frac{\partial L_k}
{\partial \boldsymbol{w}_k}= \boldsymbol{w}_k + \mu e_k
\boldsymbol{x}_k \label{eq:stndrdlmsupdate}
\end{equation}
where $\mu$ is the step size of the iterative algorithm. To make
sure that the LMS algorithm converges, $\mu$ is chosen such that
$0 < \mu <\lambda^{-1}_{\rm max}$ with $\lambda_{\rm max}$ being
the maximum eigenvalue of the covariance matrix of
$\boldsymbol{x}_k$, i.e., $\boldsymbol{R} \triangleq E \big[
\boldsymbol{x}_k \boldsymbol{x}^T_k \big]$. For the purpose of
convergence analysis of the LMS algorithm, a coefficient error
vector is usually defined as
\begin{equation}
\boldsymbol{v}_{k} \triangleq \boldsymbol{w}_k - \boldsymbol{w}.
\label{eq:coeferrvector}
\end{equation}

The data vector $\boldsymbol{x}_k$ is assumed to be independent of
the coefficient error vector $\boldsymbol{v}_k$. The excess MSE
denoted as $\xi_k$ is defined as $\xi_k \triangleq E \big[
\left(\boldsymbol{v}^T_{k} \boldsymbol{x}_k \right)^2 \big]$. It
can be further expanded as
\begin{align}
\label{eq:excessmse1} \xi_k = E \big[ \boldsymbol{v}^T_{k}
\boldsymbol{x}_k \boldsymbol{x}^T_{k} \boldsymbol{v}_k \big] .
\end{align}
In \eqref{eq:excessmse1}, $\boldsymbol{v}^T_{k} \boldsymbol{x}_k
\boldsymbol{x}^T_{k} \boldsymbol{v}_k$  is a scalar, and
therefore, it is equal to its trace, denoted hereafter as $\text{
tr} \{ \cdot \}$. Also, since $\text{ tr} \left\{
\boldsymbol{v}^T_{k} \boldsymbol{x}_k \boldsymbol{x}^T_{k}
\boldsymbol{v}_k \right\} = \text{ tr} \left\{ \boldsymbol{x}_k
\boldsymbol{x}^T_{k} \boldsymbol{v}_k \boldsymbol{v}^T_{k}
\right\}$ and the two mathematical operators of matrix trace and
expectation are interchangeable we can simplify
\eqref{eq:excessmse1} as
\begin{align}
\label{eq:excessmse} \xi_k = E \big[\text{ tr} \left\{
\boldsymbol{v}^T_{k} \boldsymbol{x}_k \boldsymbol{x}^T_{k}
\boldsymbol{v}_k \right\} \big] = \text{ tr} \left\{
\boldsymbol{R} E \big[ \boldsymbol{v}_{k} \boldsymbol{v}^T_k \big]
\right\} .
\end{align}

Let us introduce the matrix $\boldsymbol{R_v} \triangleq
\lim_{k\to\infty} E \big[ \boldsymbol{v}_{k} \boldsymbol{v}^T_{k}
\big]$ and the vector $\xi \triangleq \lim_{k \to \infty} \xi_k$.
Then we have from \eqref{eq:excessmse} that $\xi = \text{tr}
\left\{ \boldsymbol{R} \boldsymbol{R_v} \right\}$. Moreover, the
excess MSE can be found as \cite{Chenetal09}
\begin{align}
\label{eq:MSEstandardLMS} \xi = \frac{\eta}{2 - \eta}
\sigma^2_{\rm n}
\end{align}
where
\begin{align}
\eta \triangleq \mu \text{ tr} \left\{ \boldsymbol{R} \left(
\boldsymbol{I} - \mu \boldsymbol{R} \right)^{-1} \right\} .
\end{align}

\section{Reweighted $l_1$-norm Penalized LMS Algorithm}
\label{sec:rl1penallms} 
In the standard LMS algorithm, the fact
that the cost function is convex guarantees that the gradient
descent algorithm converges to the optimum point under the
aforementioned condition on $\mu$. The standard LMS algorithm
assumes no structural information about the signal/system to be
estimated. Taking any structural information into account, one
should be able to modify the algorithm and benefit by lower
estimation error, faster convergence, or lower algorithm
complexity. In this paper, we are interested in the case when the
CIR is sparse. For a CIR $\boldsymbol{w}$ to be sparse in some
sparsity domain $\boldsymbol{\Psi}$ most of the coefficients in
the vector representation of $\boldsymbol{w}$ in this domain
$\boldsymbol{\Psi}$ should be zeros or insignificant in value.
Several sparsity-aware modifications of the standard LMS have been
introduced in the literature \cite{Chenetal09, Guetal09, ShiShi10,
Suetal12, ShiMa10, YangSobelman10, Murakamietal10, OmidSergiy11}.

The reweighted $l_1$-norm minimization for
sparse signal recovery has a better performance than the standard
$l_1$-norm minimization that is usually employed in the CS
literature \cite{CandesWakinBoyd08}. It is due to
the fact that a properly reweighted $l_1$-norm approximates the
$l_0$-pseudo-norm, which actually needs to be minimized, better
than the $l_1$-norm. Therefore, one approach to enforce the
sparsity of the solution for the sparsity-aware LMS-type
algorithms is to introduce the reweighted $l_1$-norm penalty term
in the cost function \cite{OmidSergiy11}.\footnote{ The other
approach is to use the $l_p$-pseudo-norm penalty term with $0<p<1$
which is introduced in the simulations section.} Our reweighted
$l_1$-norm penalized LMS algorithm considers a penalty term
proportional to the reweighted $l_1$-norm of the coefficient
vector. The corresponding cost function can be written as
\begin{equation}
L^{\rm rl_1}_k \triangleq (1/2) e^{2}_k + \gamma_{\rm r} \|
\boldsymbol{s}_k \boldsymbol{w}_k \|_{1} \label{eq:rl1pencost}
\end{equation}
where $\| \cdot \|_{1}$ stands for the $l_1$-norm of a vector and 
$\gamma_{\rm r}$ is the weight associated with the penalty
term and elements of the $1 \times N$ row vector
$\boldsymbol{s}_k$ are set to
\begin{equation}
[\boldsymbol{s}_k]_i = \frac{1}{\epsilon_{\rm r} + |
[\boldsymbol{w}_{k-1}]_i |}, \;\;\;\; i = 1, \hdots, N
\label{eq:l1weight}
\end{equation}
with $\epsilon_{\rm r}$ being some positive number. The update
equation can be derived by differentiating \eqref{eq:rl1pencost}
with respect to the vector of CIR coefficients and using the
gradient descent principle shown in \eqref{eq:stndrdlmsupdate}.
The resulting update equation is
\begin{equation}
\boldsymbol{w}_{k+1} = \boldsymbol{w}_k + \mu e_k \boldsymbol{x}_k
- \rho_{\rm r} \frac{\text{sgn} (\boldsymbol{w}_k)} {\epsilon_{\rm
r} + | \boldsymbol{w}_{k-1} |} \label{eq:rl1eq}
\end{equation}
where $\rho_{\rm r} = \mu \gamma_{\rm r}$ and $\text{sgn} (\cdot)$ 
is the sign function which operates on every component of the vector 
separately and it is zero for $x = 0$, $1$ for $x > 0$, and $-1$ for 
$x < 0$. The absolute value operator as well as the $\text{sgn} 
(\cdot)$ and the division operator in the last term of \eqref{eq:rl1eq} 
are all component-wise. Therefore, the $i$-th element of $\text{sgn} 
(\boldsymbol{w}_k) / \left(\epsilon_{\rm r} + | \boldsymbol{w}_{k-1} |
\right)$ is $[\text{sgn} (\boldsymbol{w}_k)]_i / \left(\epsilon_{\rm r} 
+ | [\boldsymbol{w}_{k-1}]_i |\right)$. Note that although the weight
vector $\boldsymbol{s}_k$ changes in every stage of this
sparsity-aware LMS algorithm, it does not depend on
$\boldsymbol{w}_k$, and the cost function $L^{\rm rl_1}_k$ is
convex. Therefore, the reweighted
$l_1$-norm penalized LMS algorithm is guaranteed to converge to
the global minimum under some conditions. Thus, we study the
convergence of the proposed algorithm in the next section.

\section{Convergence Study of the Reweighted $l_1$-norm Penalized
LMS Method} \label{ssec:reweightedl1penalizedlmsconvergence} The
reweighted $l_1$-norm penalized LMS algorithm follows the logic
that the penalty term resembling the $l_0$-pseudo-norm of the
coefficient vector forces the solution of the modified LMS
algorithm to be sparse. The cost function of the reweighted
$l_1$-norm penalized LMS algorithm is given in
\eqref{eq:rl1pencost}, while the update equation is given in
\eqref{eq:rl1eq}.

\subsection{Mean Convergence}
We first study the mean convergence of the reweighted $l_1$-norm
penalized LMS algorithm. The update equation for the coefficient
error vector of the $l_1$-norm penalized LMS $\boldsymbol{v}_k$
can be written as
\begin{align}
\label{eq:vkplus1fromvk4rl1} \boldsymbol{v}_{k+1} & =
\boldsymbol{v}_{k} + \mu \big( (\boldsymbol{w}^T -
\boldsymbol{w}^T_k ) \boldsymbol{x}_k+n_k \big) \boldsymbol{x}_k -
\rho_{\rm r} \frac{\text{sgn} (\boldsymbol{w}_k)} {\epsilon_{\rm
r} + | \boldsymbol{w}_{k-1} |} \nonumber \\
&= \boldsymbol{v}_{k} - \mu \boldsymbol{v}^T_k \boldsymbol{x}_k
\boldsymbol{x}_k + \mu n_k \boldsymbol{x}_k - \rho_{\rm r}
\frac{\text{sgn} (\boldsymbol{w}_k)} {\epsilon_{\rm r} + |
\boldsymbol{w}_{k-1} |}.
\end{align}
Since $\boldsymbol{v}^T_k \boldsymbol{x}_k$ is a scalar which is
equal to $\boldsymbol{x}^T_k \boldsymbol{v}_k$,
\eqref{eq:vkplus1fromvk4rl1} can be rewritten as
\begin{align}
\label{eq:vkplus1fromvk4rl1final} \boldsymbol{v}_{k+1} =
\boldsymbol{v}_{k} - \mu \boldsymbol{x}_k \boldsymbol{x}^T_k
\boldsymbol{v}_k + \mu n_k \boldsymbol{x}_k - \rho_{\rm r}
\frac{\text{sgn} (\boldsymbol{w}_k)} {\epsilon_{\rm r} + |
\boldsymbol{w}_{k-1} |}.
\end{align}

From \eqref{eq:vkplus1fromvk4rl1final} we can derive the evolution
equation for $E \big[ \boldsymbol{v}_k \big]$. Since $n_k$ and
$\boldsymbol{x}_k$ are independent and $n_k$ is assumed to have
zero mean, we have $E \big[ \mu n_k \boldsymbol{x}_k \big] = 0$.
Then the evolution equation is
\begin{equation}
E \big[ \boldsymbol{v}_{k+1} \big] = \left( \boldsymbol{I} - \mu
\boldsymbol{R} \right) E \big[ \boldsymbol{v}_k \big] - \rho_{\rm
r} E \bigg[ \frac{\text{sgn} (\boldsymbol{w}_k)}{\epsilon_{\rm r}
+ | \boldsymbol{w}_{k-1} | } \bigg].
\label{eq:reweightedl1errorevolution}
\end{equation}
It is easy to see that the term $\text{sgn} (\boldsymbol{w}_k) /
(\epsilon_{\rm r} + | \boldsymbol{w}_{k-1} |)$ is bounded below 
and above element-wise as follows
\begin{equation}
\frac{-\boldsymbol{1}}{\epsilon_{\rm r}} \leq \frac{\text{sgn}
(\boldsymbol{w}_k)}{\epsilon_{\rm r} + | \boldsymbol{w}_{k-1} |}
\leq \frac{\boldsymbol{1}}{\epsilon_{\rm r}}
\label{eq:boundondgnwnminus1}
\end{equation}
where $\boldsymbol{1}$ is the vector with all of its entries set
to one. Indeed, $-1$ is always less than or equal to $\text{sgn}
(\boldsymbol{w}_k)$, while $1$ is always larger than or equal to 
$\text{sgn} (\boldsymbol{w}_k)$. Moreover, $| \boldsymbol{w}_{k-1} 
|$ and $\epsilon_{\rm r}$ are always non-negative, which means that 
the denominator of the middle term in \eqref{eq:boundondgnwnminus1} 
is always larger than or equal to the denominator of the right and 
left terms of \eqref{eq:boundondgnwnminus1}, which means that 
\eqref{eq:boundondgnwnminus1} always holds true.

We can further see that, $\rho_{\rm r} E \big[ \text{sgn}
(\boldsymbol{w}_k) / \left( \epsilon_{\rm r} +
|\boldsymbol{w}_{k-1}| \right) \big]$ is bounded between
$(-\rho_{\rm r} / \epsilon_{\rm r}) \boldsymbol{1}$ and
$(\rho_{\rm r} / \epsilon_{\rm r}) \boldsymbol{1}$. This bound on
the second term on the right hand side of
\eqref{eq:boundondgnwnminus1} is helpful for studying the mean
convergence of the reweighted $l_1$-norm penalized LMS algorithm.
The following theorem establishes our main result on the mean
convergence of the reweighted $l_1$-norm penalized LMS algorithm.

\newtheorem{rl1meanconvergence}{Theorem}
\begin{rl1meanconvergence}
\label{thm:rl1meanconvergence} If the maximal eigenvalue of the
matrix $\boldsymbol{I} - \mu \boldsymbol{R}$ is smaller than 1,
then the mean coefficient error vector $E \big[ \boldsymbol{v}_{k}
\big]$ is bounded as $k \to \infty$.
\end{rl1meanconvergence}

%\begin{proof}
Let $\boldsymbol{Q} \boldsymbol{\Lambda} \boldsymbol{Q}^T$ be the
eigenvalue decomposition of $\boldsymbol{R}$. Equation
\eqref{eq:reweightedl1errorevolution} can be rewritten as
\begin{equation}
E\big[\boldsymbol{c}_{k+1}\big] = \left( \boldsymbol{I} - \mu
\boldsymbol{\Lambda} \right) E \big[ \boldsymbol{c}_k\big] -
\boldsymbol{w}^{\prime}_{k} \label{eq:reweightedl1errorevolution1}
\end{equation}
where
\begin{eqnarray}
\boldsymbol{c}_{k} \!\!&\triangleq&\!\! \boldsymbol{Q}^T
\boldsymbol{v}_{k} \nonumber \\
\boldsymbol{w}^{\prime}_{k} \!\!&\triangleq&\!\! \rho_{\rm r}
\boldsymbol{Q}^T E \bigg[ \frac{\text{sgn} (\boldsymbol{w}_k)
}{\epsilon_{\rm r} + |\boldsymbol{w}_{k-1} |} \bigg].
\label{eq:rl1errorevolutionvars}
\end{eqnarray}
Let also $\boldsymbol{q}$ be the vector whose $i$-th entry is the
sum of the absolute values of the elements in the $i$-th row of
the matrix $\boldsymbol{Q}^T$. The variable $q_m$ is defined as
the maximum element of the vector $\boldsymbol{q}$. The vector
$\boldsymbol{Q}^T \text{sgn}(\boldsymbol{w}_k)$ is thus bounded
between $q_m \boldsymbol{1}$ and $-q_m \boldsymbol{1}$. Therefore,
the variable $\boldsymbol{w}^{\prime}_{k}$ in
\eqref{eq:rl1errorevolutionvars} is bounded between $(\rho_{\rm r}
q_m / \epsilon_{\rm r}) \boldsymbol{1}$ and $(-\rho_{\rm r} q_m /
\epsilon_{\rm r}) \boldsymbol{1}$.

It is easy to see from \eqref{eq:reweightedl1errorevolution1} that
\begin{eqnarray}
E \big[ \boldsymbol{c}_{k+M} \big] \!\!&=&\!\! \left(
\boldsymbol{I} - \mu \boldsymbol{\Lambda} \right)^M E \big[
\boldsymbol{c}_k \big] \nonumber \\
&& \quad - \sum_{m=0}^{M-1}{\left( \boldsymbol{I} - \mu
\boldsymbol{\Lambda} \right)^m \boldsymbol{w}^{\prime}_{k+M-m-1}}
. \label{eq:reweightedl1errorevolution2}
\end{eqnarray}
Moreover, since $\boldsymbol{\Lambda}$ and correspondingly
$\boldsymbol{I} - \mu \boldsymbol{\Lambda}$ are diagonal matrices,
the convergence behavior of every element of the vector $E \big[
\boldsymbol{c}_{k+M} \big]$ can be studied separately.

Let $\lambda_i$ be the $i$-th diagonal element of the matrix
$\boldsymbol{\Lambda}$. From
\eqref{eq:reweightedl1errorevolution2}, we have
\begin{align}
\label{eq:reweightedl1errorevolution3} \bigg[ E \big[
\boldsymbol{c}_{k+M} \big] \bigg]_i &= \left( 1 - \mu \lambda_i
\right)^M \bigg[ E \big[ \boldsymbol{c}_k \big] \bigg]_i
\nonumber \\
&\;\;\;\; - \sum_{m=0}^{M-1}{\left( 1-\mu\lambda_i \right)^m
\bigg[ \boldsymbol{w}^{\prime}_{k+M-m-1} \bigg]_i}
\end{align}
where $[\cdot]_i$ denotes the $i$-th entry of a vector. Since the
largest eigenvalue of $\boldsymbol{I} - \mu \boldsymbol{R}$ is
smaller than 1, then all the diagonal elements $1 - \mu \lambda_i$
are smaller than 1. Also note that the $i$-th entry of the vector
$\boldsymbol{w}^{\prime}_k$ is bounded between $\rho_{\rm r} q_m /
\epsilon_{\rm r}$ and $-\rho_{\rm r} q_m / \epsilon_{\rm r}$.
Therefore, by letting $M \to \infty$, the sum on the right hand
side of \eqref{eq:reweightedl1errorevolution3} is a geometric
series with a common ratio of $1 - \mu \lambda_i$ and is bounded
between $\rho_{\rm r} q_m / (\mu \lambda_i \epsilon_{\rm r})$ and
$-\rho_{\rm r} q_m / (\mu \lambda_i \epsilon_{\rm r})$. The other
term on the right hand side of
\eqref{eq:reweightedl1errorevolution3} approaches zero as $M \to
\infty$. As a result, $\bigg[ E \big[ \boldsymbol{c}_{k+M} \big]
\bigg]_i$ as well as the whole vector $E \big[
\boldsymbol{c}_{k+M} \big]$ are bounded when $M \to \infty$. Since
according to \eqref{eq:rl1errorevolutionvars} $E \big[
\boldsymbol{c}_k \big]$ is a rotated version of $E \big[
\boldsymbol{v}_k \big]$, the coefficient error vector
$\boldsymbol{v}_k$ is also bounded in mean. Therefore, if the
largest eigenvalue of $\boldsymbol{I} - \mu \boldsymbol{R}$ is
smaller than 1, then $E \big[ \boldsymbol{v}_k \big]$ is bounded
as $k \rightarrow \infty$.
%\end{proof}

Note that the condition in Theorem~\ref{thm:rl1meanconvergence} is
the same as the mean convergence condition for the standard LMS
algorithm which has the following evolution equation for $E \big[
\boldsymbol{v}_k \big]$
\begin{equation}
E \big[ \boldsymbol{v}_{k+1} \big] = \left( \boldsymbol{I} - \mu
\boldsymbol{R} \right) E \big[ \boldsymbol{v}_k \big].
\label{eq:LMSerrorevolution}
\end{equation}

\subsection{Excess MSE}
We now turn to the excess MSE calculation for the reweighted
$l_1$-norm penalized LMS algorithm. Using the expression in
\eqref{eq:vkplus1fromvk4rl1} for $\boldsymbol{v}_{k+1}$, the
variable $\boldsymbol{v}_{k+1} \boldsymbol{v}^T_{k+1}$ can be
written as follows
\begin{align}
\label{eq:vkvkt4rl1} \setcounter{equation}{23}
\boldsymbol{v}_{k+1} \boldsymbol{v}^T_{k+1} &=
\left(\boldsymbol{v}_{k} - \mu \boldsymbol{x}_k \boldsymbol{x}^T_k
\boldsymbol{v}_k + \mu n_k \boldsymbol{x}_k- \rho_{\rm r}
\frac{\text{sgn} (\boldsymbol{w}_k)} {\epsilon_{\rm r} + |
\boldsymbol{w}_{k-1} |}\right) \nonumber \\
& \qquad \qquad \times \left(\boldsymbol{v}^T_{k} - \mu
\boldsymbol{v}^T_k\boldsymbol{x}_k  \boldsymbol{x}^T_k + \mu n_k
\boldsymbol{x}^T_k - \rho_{\rm r} \frac{\text{sgn}
(\boldsymbol{w}^T_k)} {\epsilon_{\rm r} + | \boldsymbol{w}^T_{k-1}
|}\right).
\end{align}

Expanding the right hand side of \eqref{eq:vkvkt4rl1} and then
taking expectation of the both sides results in the following
equation
\begin{align}
\setcounter{equation}{24} \label{eq:rl1penalExpvkvkt} E \big[
\boldsymbol{v}_{k+1} &\boldsymbol{v}^T_{k+1} \big] = E \big[
\boldsymbol{v}_{k} \boldsymbol{v}^T_{k} \big] - \mu \left( E \big[
\boldsymbol{v}_{k} \boldsymbol{v}^T_{k} \boldsymbol{x}_{k}
\boldsymbol{x}^T_{k} \big] + E \big[ \boldsymbol{x}_{k}
\boldsymbol{x}^T_{k} \boldsymbol{v}_{k} \boldsymbol{v}^T_{k} \big]
\right) \nonumber \\
&+ \mu^2 E \big[ n^2_k \boldsymbol{x}_{k}
\boldsymbol{x}^T_{k} \big]
+ \mu \left( E \big[ n_k \boldsymbol{v}_{k} \boldsymbol{x}^T_{k}
\big] + E \big[ n_k
\boldsymbol{x}_{k} \boldsymbol{v}^T_{k}\big] \right) \nonumber \\
& - \mu^2 \left( E \big[ n_k \boldsymbol{x}_k \boldsymbol{x}^T_k
\boldsymbol{v}_k \boldsymbol{x}^T_k \big] \!\!+ \!\!E \big[ n_k
\boldsymbol{x}_k \boldsymbol{v}^T_k \boldsymbol{x}_k
\boldsymbol{x}^T_k \big] \right) + \mu^2 E \big[ \boldsymbol{x}_k
\boldsymbol{x}^T_k \boldsymbol{v}_k \boldsymbol{v}^T_k
\boldsymbol{x}_k \boldsymbol{x}^T_k \big] \nonumber \\
& - \rho_{\rm r} \left( E \bigg[ \boldsymbol{v}_k \frac{\text{sgn}
(\boldsymbol{w}^T_k)} {\epsilon_{\rm r} + | \boldsymbol{w}^T_{k-1}
|} \bigg] + E \bigg[ \frac{\text{sgn} (\boldsymbol{w}_k)} {
\epsilon_{\rm r} + | \boldsymbol{w}_{k-1} |} \boldsymbol{v}^T_k
\bigg] \right) \nonumber \\
& + \mu \rho_{\rm r} \left( E \bigg[ \boldsymbol{x}_k
\boldsymbol{x}^T_k \boldsymbol{v}_k \frac{\text{sgn}
(\boldsymbol{w}^T_k)} {\epsilon_{\rm r} + | \boldsymbol{w}^T_{k-1}
|} \bigg] + E \bigg[ \frac{\text{sgn} (\boldsymbol{w}_k)}
{\epsilon_{\rm r} + | \boldsymbol{w}_{k-1} |} \boldsymbol{v}^T_k
\boldsymbol{x}_k \boldsymbol{x}^T_k \bigg] \right) \nonumber \\
& - \mu \rho_{\rm r} \left( E \bigg[ n_k \boldsymbol{x}_k
\frac{\text{sgn} (\boldsymbol{w}^T_k)} {\epsilon_{\rm r} + |
\boldsymbol{w}^T_{k-1} |} \bigg] + E \bigg[ n_k \frac{\text{sgn}
(\boldsymbol{w}_k)} {\epsilon_{\rm r} + | \boldsymbol{w}_{k-1}
|}\boldsymbol{x}^T_k \bigg] \right) \nonumber \\
& + \rho^2_{\rm r} \left( E \bigg[ \frac{\text{sgn}
(\boldsymbol{w}_k)} {\epsilon_{\rm r} + | \boldsymbol{w}_{k-1}
|}\frac{\text{sgn} (\boldsymbol{w}^T_k)} {\epsilon_{\rm r} + |
\boldsymbol{w}^T_{k-1} |}\bigg] \right).
\end{align}

It is worth noting that due to the independence of the additive 
noise $n_k$ of the data and coefficient error vectors and due to 
the fact that the additive noise is zero mean, we have
\begin{align}
E \big[ n_k \boldsymbol{v}_{k} \boldsymbol{x}^T_{k}\big] &= E
\big[ n_k \big] E \big[ \boldsymbol{v}_{k} \boldsymbol{x}^T_{k}
\big] = 0 \nonumber \\
E \big[ n_k \boldsymbol{x}_{k} \boldsymbol{v}^T_{k} \big] &= E
\big[ n_k \big] E \big[ \boldsymbol{x}_{k} \boldsymbol{v}^T_{k}
\big] = 0 \nonumber \\
E \big[ n_k \boldsymbol{x}_k \boldsymbol{x}^T_k \boldsymbol{v}_k
\boldsymbol{x}^T_k \big] &= E \big[ n_k \big] E \big[
\boldsymbol{x}_k \boldsymbol{x}^T_k \boldsymbol{v}_k
\boldsymbol{x}^T_k \big] = 0 \nonumber \\
E \big[ n_k \boldsymbol{x}_k \boldsymbol{v}^T_k \boldsymbol{x}_k
\boldsymbol{x}^T_k \big] &= E \big[ n_k \big] E \big[
\boldsymbol{x}_k \boldsymbol{v}^T_k \boldsymbol{x}_k
\boldsymbol{x}^T_k \big] = 0 \nonumber \\
E \bigg[ n_k \boldsymbol{x}_k \frac{\text{sgn}
(\boldsymbol{w}^T_k)} {\epsilon_{\rm r} + | \boldsymbol{w}^T_{k-1}
|} \bigg] &= E \bigg[ n_k \frac{\text{sgn} (\boldsymbol{w}_k)}
{\epsilon_{\rm r} + | \boldsymbol{w}_{k-1} |} \boldsymbol{x}^T_k
\bigg] = 0. \nonumber
\end{align}

Since for Gaussian input sequences $E \big[ \boldsymbol{x}_k 
\boldsymbol{x}^T_k \boldsymbol{v}_k \boldsymbol{v}^T_k 
\boldsymbol{x}_k \boldsymbol{x}^T_k \big]$ can be shown to be equal 
to $2 \boldsymbol{R} E \big[ \boldsymbol{v}_{k} \boldsymbol{v}^T_{k} 
\big] \boldsymbol{R} + \boldsymbol{R} \text{ tr} \left\{ \boldsymbol{R} 
E \big[ \boldsymbol{v}_{k} \boldsymbol{v}^T_{k} \big] \right\}$ (see, 
for example, equation (12) of \cite{Douglas95} and the derivation of 
equation (35) in \cite{Horowitz}) in \eqref{eq:rl1penalExpvkvkt}, the 
expression for $E \big[ \boldsymbol{v}_{k+1} \boldsymbol{v}^T_{k+1} 
\big]$ can be derived as in the following equation
\begin{align}
\setcounter{equation}{25} \label{eq:rl1penalExpvkvkt2} E \big[
\boldsymbol{v}_{k+1} &\boldsymbol{v}^T_{k+1} \big] = E \big[
\boldsymbol{v}_{k} \boldsymbol{v}^T_{k}\big]- \mu \left( E \big[
\boldsymbol{v}_{k} \boldsymbol{v}^T_{k} \big] \boldsymbol{R} +
\boldsymbol{R} E \big[ \boldsymbol{v}_{k} \boldsymbol{v}^T_{k}
\big] \right) + \mu^2 \sigma^2_{\rm n} \boldsymbol{R} \nonumber \\
& + \mu^2 \left( 2 \boldsymbol{R} E \big[ \boldsymbol{v}_{k}
\boldsymbol{v}^T_{k} \big] \boldsymbol{R} + \boldsymbol{R} \text{
tr} \left\{ \boldsymbol{R} E \big[ \boldsymbol{v}_{k}
\boldsymbol{v}^T_{k} \big] \right\} \right) \nonumber \\
& - \rho_{\rm r} \left( ( \boldsymbol{I} - \mu \boldsymbol{R} ) E
\bigg[ \boldsymbol{v}_k \frac{\text{sgn} (\boldsymbol{w}^T_k)}
{\epsilon_{\rm r} + | \boldsymbol{w}^T_{k-1} |} \bigg] + E \bigg[
\frac{\text{sgn} (\boldsymbol{w}_k)} {\epsilon_{\rm r} + |
\boldsymbol{w}_{k-1} |} \boldsymbol{v}^T_k \bigg] ( \boldsymbol{I}
- \mu \boldsymbol{R}) \right) \nonumber \\
& + \rho^2_{\rm r} \left( E \bigg[ \frac{\text{sgn}
(\boldsymbol{w}_k)} {\epsilon_{\rm r} + | \boldsymbol{w}_{k-1} |}
\frac{\text{sgn} ( \boldsymbol{w}^T_k )} {\epsilon_{\rm r} + |
\boldsymbol{w}^T_{k-1} |} \bigg] \right).
\end{align}

Let $\boldsymbol{A}_k$ and $\boldsymbol{B}_k$ be defined as
\begin{align}
\boldsymbol{A}_k & \triangleq \rho_{\rm r} \left( (\boldsymbol{I}
- \mu \boldsymbol{R} ) E \bigg[ \boldsymbol{v}_k \frac{\text{sgn}
( \boldsymbol{w}^T_k )} {\epsilon_{\rm r} + |
\boldsymbol{w}^T_{k-1} |} \bigg]\right. \nonumber \\
& \quad\quad\quad + E \left. \bigg[ \frac{\text{sgn}
(\boldsymbol{w}_k)} {\epsilon_{\rm r} + | \boldsymbol{w}_{k-1} |}
\boldsymbol{v}^T_k \bigg] (\boldsymbol{I} - \mu \boldsymbol{R})
\right)
\end{align}
and
\begin{align}
\boldsymbol{B}_k  \triangleq \rho^2_{\rm r} \left( E \bigg[
\frac{\text{sgn} (\boldsymbol{w}_k)} {\epsilon_{\rm r} + |
\boldsymbol{w}_{k-1} |} \frac{\text{sgn} (\boldsymbol{w}^T_k)}
{\epsilon_{\rm r} + | \boldsymbol{w}^T_{k-1} |} \bigg] \right).
\end{align}
Then, \eqref{eq:rl1penalExpvkvkt2} can be rewritten as
\begin{align}
\label{eq:rl1penalExpvkvkt3} E & \big[ \boldsymbol{v}_{k+1}
\boldsymbol{v}^T_{k+1} \big] = E \big[ \boldsymbol{v}_{k}
\boldsymbol{v}^T_{k} \big] \nonumber \\
& - \mu \left( E \big[ \boldsymbol{v}_{k} \boldsymbol{v}^T_{k}
\big] \boldsymbol{R} + \boldsymbol{R} E \big[ \boldsymbol{v}_{k}
\boldsymbol{v}^T_{k} \big] \right) + \mu^2 \sigma^2_{\rm n}
\boldsymbol{R} \nonumber \\
& + \mu^2 \left( 2 \boldsymbol{R} E \big[ \boldsymbol{v}_{k}
\boldsymbol{v}^T_{k} \big] \boldsymbol{R} + \boldsymbol{R} \text{
tr} \left\{ \boldsymbol{R} E \big[ \boldsymbol{v}_{k}
\boldsymbol{v}^T_{k} \big] \right\} \right)
\nonumber \\
& - \boldsymbol{A}_k + \boldsymbol{B}_k.
\end{align}

Letting $k\to\infty$ in \eqref{eq:rl1penalExpvkvkt3}, we obtain
\begin{align}
\label{eq:rl1penalrvr1} \boldsymbol{R_v} &= \boldsymbol{R_v} - \mu
\left( \boldsymbol{R_v} \boldsymbol{R} + \boldsymbol{R}
\boldsymbol{R_v} \right) + \mu^2 \sigma^2_{\rm n} \boldsymbol{R}
\nonumber \\
&\!\!\!\!\!\!\!\!\!\!\!\!\!\!\! + \mu^2 \left( 2 \boldsymbol{R}
\boldsymbol{R_v} \boldsymbol{R} + \boldsymbol{R} \text{ tr}
\left\{ \boldsymbol{R} \boldsymbol{R_v} \right\} \right) + \lim_{k
\to \infty} \left( \boldsymbol{B}_k - \boldsymbol{A}_k \right).
\end{align}
Crossing out $\boldsymbol{R_v}$ from the both sides of
\eqref{eq:rl1penalrvr1} and then dividing the resulting equation
by $\mu$, we find that
\begin{align}
\label{eq:rl1penalrvr2} \boldsymbol{R_v} \boldsymbol{R} &
+\boldsymbol{R} \boldsymbol{R_v} - 2 \mu \boldsymbol{R}
\boldsymbol{R_v} \boldsymbol{R} \nonumber \\
& \;\;\; = \mu \boldsymbol{R} \left( \sigma^2_{\rm n} + \text{ tr}
\left\{ \boldsymbol{R} \boldsymbol{R_v} \right\} \right) +
\frac{1}{\mu} \lim_{k\to\infty} \left( \boldsymbol{B}_k -
\boldsymbol{A}_k \right).
\end{align}
Breaking $2 \mu \boldsymbol{R} \boldsymbol{R_v} \boldsymbol{R}$
into the sum of two identical terms and then factoring out
$\boldsymbol{R} \boldsymbol{R_v}$ and $\boldsymbol{R_v}
\boldsymbol{R}$, we also obtain
\begin{align}
\label{eq:rl1penalrvr3} \boldsymbol{R} \boldsymbol{R_v} &\left(
\boldsymbol{I} - \mu \boldsymbol{R} \right) + \left(
\boldsymbol{I} - \mu \boldsymbol{R} \right) \boldsymbol{R_v}
\boldsymbol{R} \nonumber \\
& \;\;\; = \mu \boldsymbol{R} \left( \sigma^2_{\rm n} + \text{ tr}
\left\{ \boldsymbol{R} \boldsymbol{R_v} \right\} \right) +
\frac{1}{\mu} \lim_{k\to\infty} \left( \boldsymbol{B}_k -
\boldsymbol{A}_k \right) .
\end{align}
Multiplying both sides of \eqref{eq:rl1penalrvr3} by $\left(
\boldsymbol{I} - \mu \boldsymbol{R} \right)^{-1}$ from right, the
following can be derived
\begin{align}
\label{eq:rl1penalrvr4} \boldsymbol{R} \boldsymbol{R_v} & + \left(
\boldsymbol{I} - \mu \boldsymbol{R} \right) \boldsymbol{R_v}
\boldsymbol{R} \left( \boldsymbol{I} - \mu \boldsymbol{R}
\right)^{-1} \nonumber \\
&\;\;\; = \mu \boldsymbol{R} \left( \boldsymbol{I} - \mu
\boldsymbol{R} \right)^{-1} \left( \sigma^2_{\rm n} + \text{ tr}
\left\{ \boldsymbol{R} \boldsymbol{R_v} \right\} \right) \nonumber
\\
& \;\;\; + \frac{1}{\mu} \lim_{k\to\infty} \left( \boldsymbol{B}_k
- \boldsymbol{A}_k \right) \left( \boldsymbol{I} - \mu
\boldsymbol{R} \right)^{-1}.
\end{align}
Note that $\sigma^2_{\rm n} + \text{ tr} \left\{ \boldsymbol{R}
\boldsymbol{R_v} \right\}$ here is a scalar. Taking the trace of
the two sides of \eqref{eq:rl1penalrvr4}, we have
\begin{align}
\label{eq:rl1penalrvr5} \text{tr} \left\{ \boldsymbol{R}
\boldsymbol{R_v} \right\} &+ \text{tr} \left\{ \left(
\boldsymbol{I} - \mu \boldsymbol{R} \right) \boldsymbol{R_v}
\boldsymbol{R} \left( \boldsymbol{I} - \mu \boldsymbol{R}
\right)^{-1} \right\} \nonumber \\
& \;\;\; = \mu \left( \sigma^2_{\rm n} + \text{ tr} \left\{
\boldsymbol{R} \boldsymbol{R_v} \right\} \right) \text{ tr}
\left\{ \boldsymbol{R} \left( \boldsymbol{I} - \mu \boldsymbol{R}
\right)^{-1} \right\} \nonumber \\
& \;\;\; + \frac{1}{\mu} \lim_{k\to\infty} \text{ tr} \left\{
\left( \boldsymbol{B}_k - \boldsymbol{A}_k \right) \left(
\boldsymbol{I} - \mu \boldsymbol{R} \right)^{-1} \right\} .
\end{align}
$\text{tr} \left\{ \left( \boldsymbol{I} - \mu
\boldsymbol{R} \right) \boldsymbol{R_v} \boldsymbol{R} \left(
\boldsymbol{I} - \mu \boldsymbol{R} \right)^{-1} \right\}$ equals 
$\text{tr} \left\{ \boldsymbol{R_v} \boldsymbol{R} \left(
\boldsymbol{I} - \mu \boldsymbol{R} \right) \left( \boldsymbol{I}
- \mu \boldsymbol{R} \right)^{-1} \right\}$ which in turn is equal
to $\text{tr} \left\{ \boldsymbol{R_v} \boldsymbol{R} \right\}$.
Therefore, equation \eqref{eq:rl1penalrvr5} can be simplified as
follows
\begin{align}
\label{eq:rl1penalrvr6} \text{tr} \left\{ \boldsymbol{R}
\boldsymbol{R_v} \right\} &+ \text{tr} \left\{ \boldsymbol{R_v}
\boldsymbol{R} \right\} \nonumber \\
& \;\;\; = \mu \left( \sigma^2_{\rm n} + \text{ tr} \left\{
\boldsymbol{R} \boldsymbol{R_v} \right\} \right) \text{ tr}
\left\{ \boldsymbol{R} \left( \boldsymbol{I} - \mu \boldsymbol{R}
\right)^{-1} \right\} \nonumber \\
& \;\;\; + \frac{1}{\mu} \lim_{k\to\infty} \text{ tr} \left\{
\left( \boldsymbol{B}_k - \boldsymbol{A}_k \right) \left(
\boldsymbol{I} - \mu \boldsymbol{R} \right)^{-1} \right\} .
\end{align}
Since $\text{tr} \left\{ \boldsymbol{R} \boldsymbol{R_v} \right\}
= \text{tr} \left\{ \boldsymbol{R_v} \boldsymbol{R} \right\}$, we
can further rewrite \eqref{eq:rl1penalrvr6} as
\begin{align}
\label{eq:rl1penalrvr7} \text{ tr} & \left\{ \boldsymbol{R}
\boldsymbol{R_v} \right\} \left( 2 - \mu \text{ tr} \left\{
\boldsymbol{R} \left( \boldsymbol{I} - \mu \boldsymbol{R}
\right)^{-1} \right\} \right) \nonumber \\
& \;\;\; = \mu \sigma^2_{\rm n} \text{ tr} \left\{ \boldsymbol{R}
\left( \boldsymbol{I} - \mu \boldsymbol{R} \right)^{-1} \right\}
\nonumber \\
& \;\;\; + \frac{1}{\mu} \lim_{k\to\infty} \text{ tr} \left\{
\left( \boldsymbol{B}_k - \boldsymbol{A}_k \right) \left(
\boldsymbol{I} - \mu \boldsymbol{R} \right)^{-1} \right\} .
\end{align}

Having in mind that the excess MSE $\xi$ is found to be $\xi =
\text{tr} \left\{ \boldsymbol{R} \boldsymbol{R_v} \right\}$, we
obtain from \eqref{eq:rl1penalrvr7} the following expression for
$\xi$:
\begin{align}
\label{eq:rl1penalxi} \xi = \text{ tr} \left\{ \boldsymbol{R}
\boldsymbol{R_v} \right\} = \frac{\eta}{2 - \eta} \sigma^2_{\rm n}
+ \frac{\beta - \alpha}{\mu (2 - \eta)}
\end{align}
where $\eta \triangleq \mu \text{tr} \left\{ \boldsymbol{R}
\left(\boldsymbol{I} - \mu\boldsymbol{R} \right)^{-1} \right\}$,
$\beta \triangleq \lim_{k\to\infty} \beta_k$, $\alpha \triangleq
\lim_{k \to \infty} \alpha_k$, $\beta_k \triangleq \text{tr}
\left\{ \boldsymbol{B}_k \left( \boldsymbol{I} - \mu
\boldsymbol{R} \right)^{-1} \right\}$, and $\alpha_k \triangleq
\text{tr} \left\{ \boldsymbol{A}_k \left( \boldsymbol{I} - \mu
\boldsymbol{R} \right)^{-1} \right\}$.

We now further examine variables $\beta_k$ and $\alpha_k$. The
matrix $\boldsymbol{B}_k \left( \boldsymbol{I} - \mu
\boldsymbol{R} \right)^{-1}$ can be expressed as
\begin{align}
\label{eq:BtimesIminusMuR} &\boldsymbol{B}_k \left( \boldsymbol{I}
- \mu \boldsymbol{R} \right)^{-1} \nonumber \\
& \quad = \rho^2_{\rm r} \left( E \bigg[ \frac{\text{sgn}
(\boldsymbol{w}_k)} {\epsilon_{\rm r} + | \boldsymbol{w}_{k-1} |}
\frac{\text{sgn} (\boldsymbol{w}^T_k)} {\epsilon_{\rm r} + |
\boldsymbol{w}^T_{k-1} |} \bigg] \left( \boldsymbol{I} - \mu
\boldsymbol{R} \right)^{-1} \right).
\end{align}
Using \eqref{eq:BtimesIminusMuR}, we obtain
\begin{align}
\label{eq:trBtimesIminusMuR1} &\beta_k = \text{ tr} \left\{
\boldsymbol{B}_k \left( \boldsymbol{I} - \mu \boldsymbol{R}
\right)^{-1}\right\} \nonumber \\
& = \rho^2_{\rm r} \left( E \bigg[ \text{ tr} \left\{
\frac{\text{sgn} (\boldsymbol{w}_k)} {\epsilon_{\rm r} + |
\boldsymbol{w}_{k-1} |} \frac{\text{sgn} (\boldsymbol{w}^T_k)}
{\epsilon_{\rm r} + | \boldsymbol{w}^T_{k-1} |} \left(
\boldsymbol{I} - \mu \boldsymbol{R} \right)^{-1} \right\} \bigg]
\right).
\end{align}
Moreover, $\beta_k$ in \eqref{eq:trBtimesIminusMuR1} can also be
written as
\begin{align}
\label{eq:trBtimesIminusMuR} \beta_k \!=\! \rho^2_{\rm r} \!
\left( \! E \bigg[ \text{tr} \left\{ \frac{\text{sgn}
(\boldsymbol{w}^T_k)} {\epsilon_{\rm r} \!+\! |
\boldsymbol{w}^T_{k-1} |} \left( \boldsymbol{I} - \mu
\boldsymbol{R} \right)^{-1} \frac{\text{sgn} (\boldsymbol{w}_k)}
{\epsilon_{\rm r} \!+\! | \boldsymbol{w}_{k-1} |} \right\}
\!\bigg]\! \right) .
\end{align}

The matrix $\boldsymbol{I} - \mu\boldsymbol{R}$ is symmetric, and
its eigenvalue decomposition can be written as $\boldsymbol{I} -
\mu \boldsymbol{R} = \boldsymbol{U} \boldsymbol{\Gamma}
\boldsymbol{U}^T$ with $\boldsymbol{U}$ being an orthonormal
matrix of eigenvectors and $\boldsymbol{\Gamma}$ being a diagonal
matrix of eigenvalues. Therefore, $\left( \boldsymbol{I} - \mu
\boldsymbol{R} \right)^{-1} = \boldsymbol{U}
\boldsymbol{\Gamma}^{-1} \boldsymbol{U}^T$ and $\beta_k$ from
equation \eqref{eq:trBtimesIminusMuR} can be written as
\begin{align}
\label{eq:betakInequality1} \beta_k &= \rho^2_{\rm r} \left( E
\bigg[ \text{tr} \left\{ \frac{\text{sgn} (\boldsymbol{w}^T_k)}
{\epsilon_{\rm r} + | \boldsymbol{w}^T_{k-1} |} \boldsymbol{U}
\boldsymbol{\Gamma}^{-1} \boldsymbol{U}^T \frac{\text{sgn}
(\boldsymbol{w}_k)} {\epsilon_{\rm r} + | \boldsymbol{w}_{k-1} |}
\right\} \bigg] \right) \nonumber \\
& = \rho^2_{\rm r} \left( E \bigg[ \text{tr} \left\{
\boldsymbol{\Gamma}^{-1} \boldsymbol{U}^T \frac{\text{sgn}
(\boldsymbol{w}_k)} {\epsilon_{\rm r} + | \boldsymbol{w}_{k-1} |}
\frac{\text{sgn} (\boldsymbol{w}^T_k)} {\epsilon_{\rm r} + |
\boldsymbol{w}^T_{k-1} |} \boldsymbol{U} \right\} \bigg] \right) .
\end{align}

Let $\lambda_{max}$ be the largest eigenvalue of the covariance
matrix $\boldsymbol{R}$. Also, let $\mu$ be small enough such that
$\left( 1 - \mu \lambda_{max} \right)^{-1}$ is positive. In
\eqref{eq:betakInequality1}, since $\boldsymbol{\Gamma}^{-1}$ is a
diagonal matrix whose diagonal elements are all non-negative and
less than or equal to $\left( 1 - \mu \lambda_{max} \right)^{-1}$,
we have
\begin{align}
\label{eq:betakInequality2} \beta_k & \leq \frac{\rho^2_{\rm r}}{1
- \mu \lambda_{max}} \nonumber \\
& \quad \times \left( E \bigg[ \text{tr} \left\{ \boldsymbol{U}^T
\frac{\text{sgn} (\boldsymbol{w}_k)} {\epsilon_{\rm r} + |
\boldsymbol{w}_{k-1} |} \frac{\text{sgn} (\boldsymbol{w}^T_k)}
{\epsilon_{\rm r} + | \boldsymbol{w}^T_{k-1} |} \boldsymbol{U}
\right\}\bigg] \right).
\end{align}
Note that
\begin{align}
\label{eq:betakInequalityaux1} &\text{tr} \left\{ \boldsymbol{U}^T
\frac{\text{sgn} (\boldsymbol{w}_k)} {\epsilon_{\rm r} + |
\boldsymbol{w}_{k-1} |} \frac{\text{sgn} (\boldsymbol{w}^T_k)}
{\epsilon_{\rm r} + | \boldsymbol{w}^T_{k-1} |} \boldsymbol{U}
\right\} \nonumber \\
& \quad = \text{tr} \left\{ \frac{\text{sgn} (\boldsymbol{w}^T_k)}
{ \epsilon_{\rm r} + | \boldsymbol{w}^T_{k-1} |} \boldsymbol{U}
\boldsymbol{U}^T \frac{\text{sgn} (\boldsymbol{w}_k)} {
\epsilon_{\rm r} + | \boldsymbol{w}_{k-1} |} \right\} \nonumber \\
& \qquad =\frac{\text{sgn} (\boldsymbol{w}^T_k)} {\epsilon_{\rm r}
+ | \boldsymbol{w}^T_{k-1} |} \frac{\text{sgn} (\boldsymbol{w}_k)}
{\epsilon_{\rm r} + | \boldsymbol{w}_{k-1} |} \nonumber \\
& \qquad \quad \leq \frac{\text{sgn} (\boldsymbol{w}^T_k)
\text{sgn} (\boldsymbol{w}_k)}{\epsilon^2_{\rm r}} \leq
\frac{N}{\epsilon^2_{\rm r}} .
\end{align}

Substituting \eqref{eq:betakInequalityaux1} in
\eqref{eq:betakInequality2}, the following bound on $\beta_k$ can
be finally obtained
\begin{align}
\label{eq:betakInequality} \beta_k \leq \frac{N\rho^2_{\rm
r}}{\epsilon^2_{\rm r} \left( 1 - \mu \lambda_{max} \right)} .
\end{align}
Moreover, $\beta_k$ in \eqref{eq:betakInequality1} can also be
written as
\begin{align}
\label{eq:betakInequality3} \beta_k = \rho^2_{\rm r} \left( E
\bigg[ \text{tr} \left\{ \boldsymbol{z}^T_k \boldsymbol{z}_k
\right\} \bigg] \right) = \rho^2_{\rm r} \left( E \bigg[ \|
\boldsymbol{z}_k \|^2_{2} \bigg] \right)
\end{align}
where $\boldsymbol{z}_k$ is defined as
\begin{align}
\label{eq:betakInequalityzk} \boldsymbol{z}_k \triangleq
\boldsymbol{\Gamma}^{-1/2} \boldsymbol{U}^T \frac{\text{sgn}
(\boldsymbol{w}_k)} {\epsilon_{\rm r} + | \boldsymbol{w}_{k-1} |}
\end{align}
and $\|\cdot\|_{2}$ stands for the Euclidean norm of a vector.
Therefore, it can be seen from \eqref{eq:betakInequality3} that
$\beta_k$ is non-negative. Since, $\beta_k$ is upper bounded and
non-negative, so is $\beta$.

The variable $\alpha_k$ can be derived as 
\begin{align}
\label{eq:trAtimesIminusMuR} \setcounter{equation}{46} \alpha_k &=
\text{tr} \left\{ \boldsymbol{A}_k \left( \boldsymbol{I} - \mu
\boldsymbol{R} \right)^{-1} \right\} = \rho_{\rm r} \left( E
\bigg[ \text{tr} \left\{ \boldsymbol{v}_k \frac{\text{sgn}
(\boldsymbol{w}^T_k)} {\epsilon_{\rm r} + | \boldsymbol{w}^T_{k-1}
|} + \frac{\text{sgn} (\boldsymbol{w}_k)} {\epsilon_{\rm r} + |
\boldsymbol{w}_{k-1} |} \boldsymbol{v}^T_k \right\} \bigg] \right)
\nonumber \\
& = 2 \rho_{\rm r} \left( E \bigg[ \text{tr} \left\{
\boldsymbol{v}_k \frac{\text{sgn} (\boldsymbol{w}^T_k)}
{\epsilon_{\rm r} + | \boldsymbol{w}^T_{k-1} |} \right\} \bigg]
\right) \nonumber \\
& = 2 \rho_{\rm r} \left( E \bigg[ \text{tr} \left\{
\boldsymbol{w}_k \frac{\text{sgn} (\boldsymbol{w}^T_k)}
{\epsilon_{\rm r} + | \boldsymbol{w}^T_{k-1} |} \right\} \bigg] -
E \bigg[ \text{tr} \left\{ \boldsymbol{w} \frac{\text{sgn}
(\boldsymbol{w}^T_k)} {\epsilon_{\rm r} + | \boldsymbol{w}^T_{k-1}
|} \right\} \bigg] \right) .
\end{align}

Assuming that $\lim_{k \to \infty} E \big[ \text{sgn}
(\boldsymbol{w}_k) \big] = \text{sgn} (\boldsymbol{w})$ which is a
common assumption and it is, for example, the same as in
\cite{Chenetal09}, $\alpha_k$ in \eqref{eq:trAtimesIminusMuR} can
be written as
\begin{align}
\label{eq:alfaksimplified} \alpha_k = 2 \rho_{\rm r} \left( E
\bigg[ \left\| \frac{\boldsymbol{w}_k} {\epsilon_{\rm r} + |
\boldsymbol{w}_{k-1} |} \right\|_{1} \bigg] \!\!-\!\! E \bigg[
\left\| \frac{\boldsymbol{w}} {\epsilon_{\rm r} + |
\boldsymbol{w}_{k-1} |} \right\|_{1} \bigg] \right) .
\end{align}

Defining $\beta^\prime \triangleq \beta / \rho^2_{\rm r}$, and
$\alpha^\prime \triangleq \alpha / \rho_{\rm r}$, the excess MSE
equation of \eqref{eq:rl1penalxi} can be rewritten as
\begin{align}
\label{eq:rl1penalximodified} \xi = \frac{\eta}{2 - \eta}
\sigma^2_{\rm n} + \frac{\beta^\prime \rho_{\rm r}} {\mu (2 -
\eta)} \left( \rho_{\rm r} - \frac{\alpha^\prime}{\beta^\prime}
\right)
\end{align}
where $\beta^\prime$ is non-negative and upper bounded by $N /
\epsilon^2_{\rm r} \left( 1 - \mu \lambda_{max} \right)$, and
$\alpha^\prime$ is given as
\begin{align}
\label{eq:alphaPrime} \alpha^\prime &= \lim_{k\to\infty} 2 \left(
E \bigg[ \left\| \frac{\boldsymbol{w}_k} {\epsilon_{\rm r} + |
\boldsymbol{w}_{k-1} |} \right\|_{1} \bigg] \right. \nonumber \\
&\qquad \qquad \qquad - E \left. \bigg[ \left\|
\frac{\boldsymbol{w}} {\epsilon_{\rm r} + | \boldsymbol{w}_{k-1}
|} \right\|_{1} \bigg] \right) .
\end{align}

It can be seen from \eqref{eq:rl1penalximodified} that if
$\alpha^\prime$ is positive, then choosing $\rho_{\rm r}$ in a way
that $\rho_{\rm r} < \alpha^\prime / \beta^\prime$ can lead to the
excess MSE of the reweighted $l_1$-norm penalized LMS algorithm
being smaller than that of the standard LMS algorithm given in
\eqref{eq:MSEstandardLMS}. The following example shows how the
value of $\alpha^\prime$ varies with respect to the sparsity level
of the CIR that is being estimated.

{\it Example 1:} A time sparse CIR of length $N=16$ whose sparsity
level varies from $S=1$ to $S=16$ is considered in this example.
The nonzero entries of the CIR take the values of $1$ or $-1$ with
equal probabilities each equal to half. In order to ensure a
constant value for the term $\eta \sigma^2_{\rm n} / ( 2 - \eta )$
in the excess MSE equation of \eqref{eq:rl1penalximodified} for
different values of sparsity $S$, $\sigma^2_{\rm n}$ is a constant
set to $0.01$. The step size $\mu$ is set to $0.05$, while
$\rho_{\rm r} = 5 \times 10^{-4}$ and $\epsilon_{\rm r} = 0.05$ in
\eqref{eq:rl1eq}. Elements of the training sequence
$\boldsymbol{x}_k$ are chosen with equal probability from the set
$\{1, -1\}$. Table~\ref{table:alphaPrimeExample} shows the value
of $\alpha^\prime$ after $250$ iterations of the reweighted
$l_1$-norm penalized LMS algorithm for different sparsity levels.

\begin{table}[htb]
    \caption{Value of $\alpha^\prime$ for different sparsity levels.}
\begin{center}
    \begin{tabular} {| c || c | c | c | c | c | c | c | c | c | c |}
    \hline
    $S$ & 1 & 2 & 3 & 4 & 5 & 6 & 7 & 8 \\ \hline
    $\alpha^\prime$ & \!\!3.23 & \!\!2.99 & \!\!2.74 & \!\!2.45 &
    \!\!2.11 & \!\!1.74 & \!\!1.32 & \!\!0.89 \\ \hline \hline
    $S$ & 9 & 10 & 11 & 12 & 13 & 14 & 15 & 16 \\ \hline
    $\alpha^\prime$ & \!\!0.39 & \!\!-0.17 & \!\!-0.79 & \!\!-1.46 &
    \!\!-2.23 & \!\!-3.10 & \!\!-4.07 & \!\!-5.20 \\ \hline
    \end{tabular}
    \end{center}
\label{table:alphaPrimeExample}
\end{table}

The results in Table~\ref{table:alphaPrimeExample} show that as
the CIR becomes less and less sparse, i.e., as $S$ increases,
$\alpha^\prime$ becomes smaller to a point that it takes a
negative value. Therefore, based on \eqref{eq:rl1penalximodified}
we can expect a smaller excess MSE for the reweighted $l_1$-norm
penalized LMS algorithm compared to that of the standard LMS
algorithm providing that the sparsity level is small enough so
that $\alpha^\prime$ is positive.

\section{Simulation Results}
\label{sec:SimRes} In this section we compare the performance of
different channel estimation algorithms for several scenarios. The
algorithms being considered here are the ZA-LMS and RZA-LMS
algorithms of \cite{Chenetal09} as well as the proposed reweighted
$l_1$-norm penalized LMS algorithm and the $l_p$-pseudo-norm
penalized LMS algorithm \cite{OmidSergiy11}. The standard LMS
algorithm is also included for comparison in our simulation
figures. The performance of the so-called \textit{oracle} LMS is
reported in the first simulation example as a lower bound for all
sparsity-aware algorithms. In oracle LMS, the positions of the
nonzero taps of the CIR are assumed to be known before hand.

The cost function of ZA-LMS can be written as $L^{\rm ZA}_k \triangleq 
(1/2) e^{2}_k + \gamma_{\rm ZA} \| \boldsymbol{w}_k \|_{1}$, where 
$\gamma_{\rm ZA}$ is the weight associated with the penalty term. The 
CIR is assumed to be sparse in the time domain and the cost function 
$L^{\rm ZA}_k$ is convex. The algorithm has the following update equation
\begin{equation}
\boldsymbol{w}_{k+1} =  \boldsymbol{w}_k + \mu e_k
\boldsymbol{x}_k - \rho_{\rm ZA} \text{sgn} (\boldsymbol{w}_k)
\label{eq:ZALMSeq}
\end{equation}
where $\rho_{\rm ZA} \triangleq \mu \gamma_{\rm ZA}$.

The RZA-LMS algorithm uses a logarithmic penalty term. The modified cost 
function of the algorithm is $L^{\rm RZA}_k \triangleq (1/2) e^{2}_k + 
\gamma_{\rm RZA} \sum_{i=1}^{N} \log \left( 1 + [ \boldsymbol{w}_k 
]_{i} / \epsilon_{\rm RZA}^{\prime} \right)$, where $[\boldsymbol{w}_k
]_{i}$ is the $i$-th element of the vector $\boldsymbol{w}_k$ and 
$\gamma_{\rm RZA}$ and $\epsilon_{\rm RZA}^{\prime}$ are some positive 
numbers. Note that the same penalty term is also used, for example, in 
\cite{Sacchi1}. The update equation for the RZA-LMS is
\begin{equation}
\boldsymbol{w}_{k+1} = \boldsymbol{w}_k + \mu e_k \boldsymbol{x}_k
- \rho_{\rm RZA} \frac{\text{sgn} (\boldsymbol{w}_k)}{1 +
\epsilon_{\rm RZA} |\boldsymbol{w}_k |} \label{eq:RZALMSeq}
\end{equation}
where $\rho_{\rm RZA} \triangleq \mu \gamma_{\rm RZA} \epsilon_{\rm 
RZA}$ and $\epsilon_{\rm RZA} \triangleq 1 / \epsilon_{\rm RZA}^{\prime}$. 
Note that the cost function of the RZA-LMS method is not convex that 
makes the convergence and consistency analysis problematic.

Although only time domain sparsity
is considered in \cite{Chenetal09}, the ZA-LMS algorithm, for
example, can be easily extended to an arbitrary sparsity basis.
Let $\boldsymbol{\Psi}$ be the $N \times N$ orthonormal matrix
denoting a specific sparsity basis. The CIR $\boldsymbol{w}$ is
sparse in the sparsity domain $\boldsymbol{\Psi}$ if its
representation in $\boldsymbol{\Psi}$, that is, the vector
$\boldsymbol{\Psi} \boldsymbol{w}$, has
only few nonzero components. The ZA-LMS cost function can be
rewritten then as $L^{\rm ZA}_k \triangleq (1/2) e^{2}_k +
\gamma_{\rm ZA} \| \boldsymbol{\Psi} \boldsymbol{w}_k \|_{1}$,
and the update equation becomes
\begin{equation}
\boldsymbol{w}_{k+1} = \boldsymbol{w}_k + \mu e_k \boldsymbol{x}_k
- \rho_{\rm ZA} \text{sgn} (\boldsymbol{\Psi} \boldsymbol{w}_k)
\boldsymbol{\Psi} \label{eq:ZALMSGeneralPsieq}
\end{equation}
where $\text{sgn} (\boldsymbol{\Psi} \boldsymbol{w}_k)$ as well as
$\text{sgn} (\boldsymbol{\Psi} \boldsymbol{w}_k)
\boldsymbol{\Psi}$ are row vectors.

In \cite{OmidSergiy11}, we considered the $l_p$-pseudo-norm of
$\boldsymbol{w}_k$ with $0 < p < 1$ as the penalty term introduced
into the cost function of the standard LMS. The cost function of
the $l_p$-pseudo-norm penalized LMS is then expressed as
$L^{\rm l_p}_k \triangleq (1/2) e^{2}_k + \gamma_{\rm p} \|
\boldsymbol{w}_k \|_{p} \label{eq:lppencost}$,
where $\| \cdot \|_{p}$ stands for the $l_p$-pseudo-norm of a
vector and $\gamma_{\rm p}$ is the corresponding weight term.
Using gradient descent, the update equation based on
\eqref{eq:lppencost} can be derived as
\begin{equation}
\boldsymbol{w}_{k+1} = \boldsymbol{w}_k + \mu e_k \boldsymbol{x}_k
- \rho_{\rm p} \frac{\left( \| \boldsymbol{w}_k \|_{p}
\right)^{1-p} \text{sgn} ( \boldsymbol{w}_k)}{| \boldsymbol{w}_k
|^{(1-p)}} \label{eq:lpeq}
\end{equation}
where $\rho_{\rm p} = \mu \gamma_{\rm p}$. In practice, we need to
impose an upper bound on the last term in \eqref{eq:lpeq} in the
situation when an entry of $\boldsymbol{w}_k$ approaches zero,
which is the case for a sparse CIR. Then the update equation
\eqref{eq:lpeq} is modified as
\begin{equation}
\boldsymbol{w}_{k+1} = \boldsymbol{w}_k + \mu e_k \boldsymbol{x}_k
- \rho_{\rm p} \frac{ \left( \| \boldsymbol{w}_k\|_{p}
\right)^{1-p} \text{sgn} (\boldsymbol{w}_k )} {\epsilon_{\rm p} +
| \boldsymbol{w}_k |^{(1-p)}} \label{eq:modifiedlpeq}
\end{equation}
where $\epsilon_{\rm p}$ is a value which is used to upper bound
the last term in \eqref{eq:lpeq}.

\subsection{Simulation Example 1: Time Sparse Channel Estimation}
In this example, we consider the problem of estimating a CIR of
length $N=16$. The CIR is assumed to be sparse in the time domain.
Two different sparsity levels of $S=1$ and $S=4$ are considered.
The positions of the nonzero taps in the CIR are chosen randomly.
The value of each nonzero tap is a zero mean Gaussian random
variable with a variance of 1.

Two different signal-to-noise ratio (SNR) values of $10$~dB and
$20$~dB are considered. For the $l_p$-pseudo-norm penalized LMS
algorithm, $p$ is chosen to be $1/2$ with $\epsilon_{\rm p} =
0.05$ and $\rho_{\rm p} = 2 \times 10^{-4}$. The parameters of the
reweighted $l_1$-norm penalized LMS algorithm are set to
$\rho_{\rm r} = 2 \times 10^{-4}$ and $\epsilon_{\rm r} = 0.05$.
For the ZA-LMS and the RZA-LMS algorithms, $\rho_{\rm ZA} = 5
\times10^{-4}$, $\rho_{\rm RZA} = 4 \times10^{-3}$, and
$\epsilon_{\rm RZA} = 25$. Parameter values for the ZA-LMS and
RZA-LMS algorithms are optimized through simulations. The step
size is set to $\mu=0.05$ for all algorithms. The measure of
performance is the MSE between the actual and estimated CIR.
Simulation results are averaged over $10000$ simulation runs to
smooth out the curves.

\begin{figure}%[t]
\centering
\subfigure[$10$ dB SNR] {
 \includegraphics[scale = 0.48]{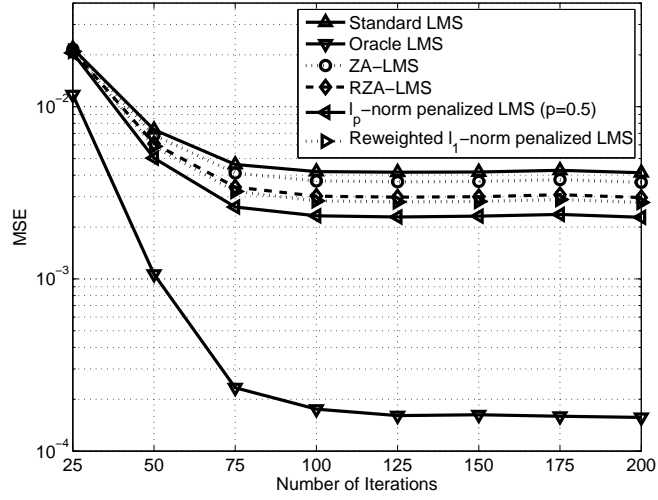}
\label{sfig:sim1case1snr10} }
\subfigure[$20$ dB SNR] {
 \includegraphics[scale = 0.48]{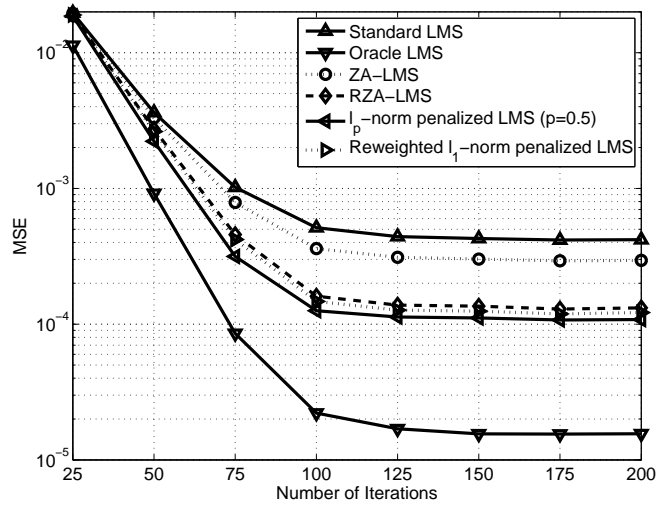}
\label{sfig:sim1case1snr20} }
\caption{Example~1, Case~1: MSE's of different estimation
algorithms vs number of iterations ($S=1$).} \label{fig:sim1case1}
\end{figure}

Fig.~\ref{fig:sim1case1} shows the MSE versus the number of
iterations for different estimation algorithms for the case when
the sparsity level is $S = 1$. It is expected that the oracle LMS
outperforms all sparsity-aware algorithms as well as the standard
LMS. The simulation results conform it. Outside the oracle LMS, it
can be seen that for both SNR values tested, the $l_p$-pseudo-norm
penalized LMS algorithm has the best performance followed by the
reweighted $l_1$-norm penalized LMS algorithm, and then by the
RZA-LMS, ZA-LMS, and standard LMS algorithms. The MSEs of the
RZA-LMS and reweighted $l_1$-norm penalized LMS algorithms are
close to each other. As the SNR increases, the performance of all
the algorithms tested improves as expected. Also, it can be seen
in Fig.~\ref{fig:sim1case1} that the performance gap between the
MSE of the standard LMS algorithm and the MSE's for the rest of
the algorithms increases as SNR increases. The $l_p$-pseudo-norm
penalized LMS and reweighted $l_1$-norm penalized LMS algorithms
have faster convergence rate compared to the standard LMS
algorithm.

\begin{figure}%[t]
\centering
\subfigure[$10$ dB SNR] {
 \includegraphics[scale = 0.48]{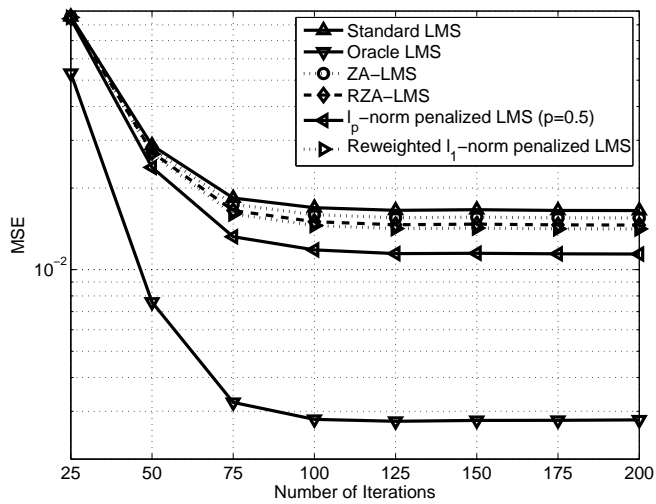}
\label{sfig:sim1case2snr10} }
\subfigure[$20$ dB SNR] {
 \includegraphics[scale = 0.48]{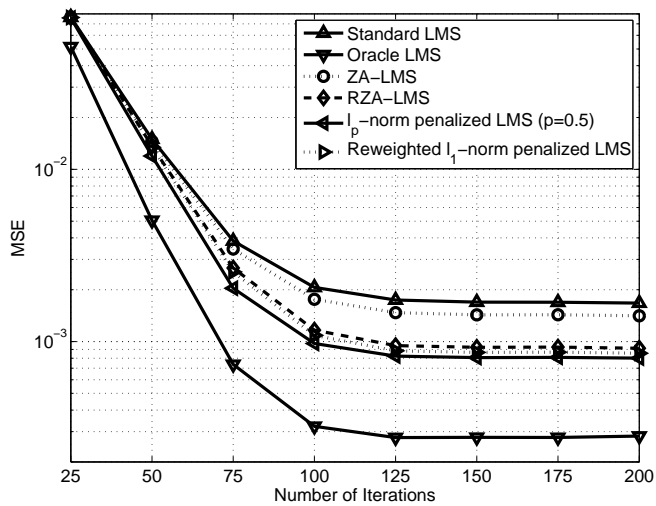}
\label{sfig:sim1case2snr20} }
\caption{Example~1, Case~2: MSE's of different estimation
algorithms vs number of iterations ($S=4$).} \label{fig:sim1case2}
\end{figure}

Fig.~\ref{fig:sim1case2} shows the simulation results for the case
when the sparsity level is set to $S = 4$. The parameter choices
for all the algorithms tested are the same as in the previous
case. Most of the observations from Fig.~\ref{fig:sim1case1} also
hold for this case of increased sparsity level. However,
increasing the sparsity level of the CIR leads to a decrease in
the performance gap between the sparsity-aware LMS algorithms and
the standard LMS algorithm.

Overall, the proposed reweighted $l_1$-norm penalized LMS algorithm 
performs better than the RZA-LMS and significantly better than the 
RA-LMS. Both the proposed reweighted $l_1$-norm penalizes LMS and 
the RA-LMS algorithms use $l_1$-norm penalty for enforcing sparsity, 
but the proposed algorithm uses the reweighting on the top. Thus, 
the corresponding performance improvement of the proposed algorithm 
as compared to the RA-LMS algorithm is due to the reweighting only. 
The RZA-LMS algorithm uses a different nonconvex penalty term, and 
it is proper to compare it to the other proposed $l_p$-pseudo-norm 
($p < 1$) penalized LMS algorithm, where the penalty term is also 
nonconvex. We can see the significant performance improvement for 
the other proposed algorithm versus the RZA-LMS algorithm.

\subsection{Simulation Example 2: Arbitrary Sparsity Basis}
The ZA-LMS and RZA-LMS algorithms in the form derived in
\cite{Chenetal09} are only applied to the case when the channel is
sparse in the time domain. However, these algorithms as well as
the $l_p$-pseudo-norm penalized LMS and reweighted $l_1$-norm
penalized LMS algorithms can be modified to accommodate the case
of an arbitrary sparsity basis. Consider the ZA-LMS algorithm in
the case when the CIR is sparse in a sparsity domain denoted by
$\boldsymbol{\Psi}$. The CIR representation in
$\boldsymbol{\Psi}$, i.e., the vector ${\boldsymbol{w}}_{\Psi} =
\boldsymbol{\Psi} \boldsymbol{w}$, is a sparse vector and it has a
few nonzero entries. The corresponding update equation for the
ZA-LMS algorithm is given by \eqref{eq:ZALMSGeneralPsieq}.

The update equation for the reweighted $l_1$-norm penalized LMS
algorithm becomes
\begin{equation}
\boldsymbol{w}_{k+1} = \boldsymbol{w}_k + \mu e_k \boldsymbol{x}_k
- \rho_{\rm r} \frac{\text{sgn} (\boldsymbol{\Psi}
\boldsymbol{w}_k) \boldsymbol{\Psi}} {\epsilon_{\rm r} + |
\boldsymbol{\Psi} \boldsymbol{w}_{k-1} |}.
\label{eq:rl1eqGeneralPsi}
\end{equation}

Finally, the modified update equation of the $l_p$-pseudo-norm
penalized LMS algorithm can be derived as
\begin{equation}
\boldsymbol{w}_{k+1} = \boldsymbol{w}_k + \mu e_k \boldsymbol{x}_k
- \rho_{\rm p} \frac{ \left( \| \boldsymbol{\Psi} \boldsymbol{w}_k
\|_{p} \right)^{1-p} \text{sgn} (\boldsymbol{\Psi}
\boldsymbol{w}_k ) \boldsymbol{\Psi}} {\epsilon_{\rm p} + |
\boldsymbol{\Psi} \boldsymbol{w}_k |^{(1-p)}}.
\label{eq:modifiedlpGeneralPsi}
\end{equation}

In this simulation example, a CIR of length $N=16$ with the
sparsity level of $S=2$ is being estimated which is sparse in the
discrete cosine transform (DCT) domain. The positions of nonzero
taps in the DCT domain are chosen randomly. The value of the
nonzero elements in the DCT domain are set to $1$ or $-1$ with the
same probabilities each equal to half. The algorithms being
compared here are the ZA-LMS, RZA-LMS, $l_p$-pseudo-norm penalized
LMS, reweighted $l_1$-norm penalized LMS, and standard LMS
algorithms. As in the first simulation scenario, two different SNR
values of $10$ and $20$~dBs are tested. Parameter choices for the
$10$~dB SNR case are as follows. For the $l_p$-pseudo-norm
penalized LMS algorithm, $p = 1/2$, $\epsilon_{\rm p} = 0.05$, and
$\rho_{\rm p} = 2 \times 10^{-4}$. Parameters of the reweighted
$l_1$-norm penalized LMS algorithm are $\rho_{\rm r} = 2 \times
10^{-4}$ and $\epsilon_{\rm r} = 0.05$. For the ZA-LMS and the
RZA-LMS algorithms, the values are $\rho_{\rm ZA} = 5
\times10^{-4}$, $\rho_{\rm RZA} = 4 \times10^{-3}$, and
$\epsilon_{\rm RZA} = 25$. The step size $\mu$ is set to $0.05$.
For the $20$~dB SNR case, $\rho_{\rm r}$, $\rho_{\rm p}$, and
$\rho_{\rm RZA}$ are reduced by half.

\begin{figure}%[t]
\centering
\includegraphics[scale = 0.48]{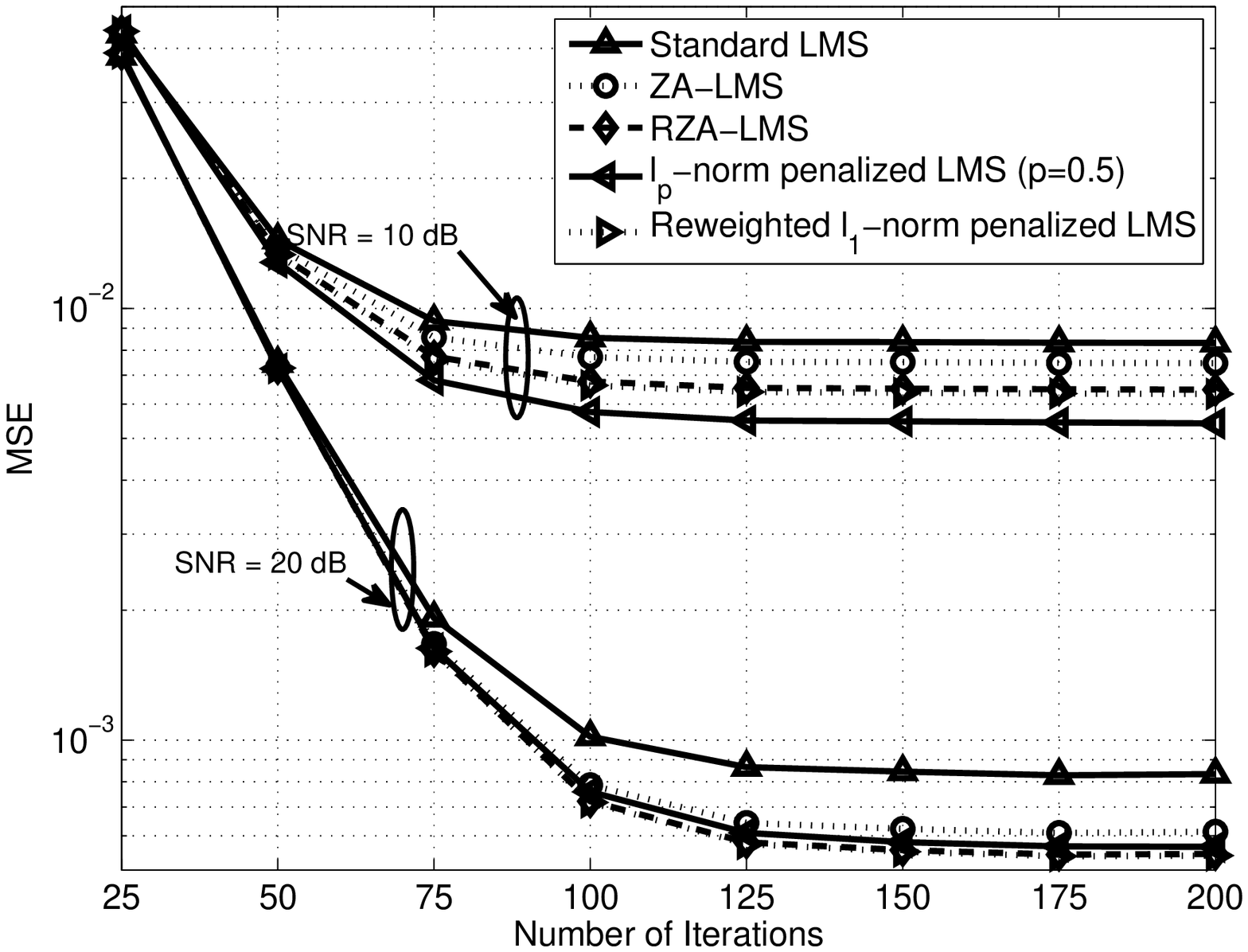}
\caption{Example 2: MSEs of estimation algorithms vs number of
iterations for a DCT sparse channel with $S=2$.} \label{fig:sim2}
\end{figure}

The MSE curves in Fig.~\ref{fig:sim2} are averaged over $10000$
simulation runs. The same conclusions as in Simulation Example~1
hold here as well. For the SNR of $10$~dB SNR, the
$l_p$-pseudo-norm penalized LMS algorithm outperforms all the
other algorithms followed by the reweighted $l_1$-norm penalized
LMS algorithm, and then by the RZA-LMS and ZA-LMS algorithms.
However, when the SNR is set to $20$~dB, the reweighted $l_1$-norm
penalized LMS and RZA-LMS algorithms show a better performance
than the $l_p$-pseudo-norm penalized LMS algorithm.

\subsection{Simulation Example 3: Effect of Sparsity Level on the
Performance of the Reweighted $l_1$-norm Penalized LMS Algorithm}
In this example, we study the effect that the increasing sparsity
level of CIR has on the performance of the reweighted $l_1$-norm
penalized LMS algorithm. A CIR is assumed to be sparse in the time
domain and it is of length $N=16$. The sparsity level varies from
$2$ to $8$. The positions of the nonzero taps of the CIR are
chosen randomly and the values of nonzero taps are set to $1$ or
$-1$ with equal probability each equal to half. Parameters of the
reweighted $l_1$-norm penalized LMS algorithm are $\rho_{\rm r} =
2 \times 10^{-4}$ and $\epsilon_{\rm r} = 0.05$. The step size
$\mu$ is set to $0.05$. Variance of the additive noise term $n_k$
is $\sigma^2_{\rm n} = 0.01$. Excess MSE is used as a performance
measure in this example. We have chosen a constant variance
$\sigma^2_{\rm n}$ for the noise in order to make sure that the
standard LMS algorithm has the same excess MSE regardless of the
sparsity level of the channel. The excess MSE curves are averaged
over $10000$ simulation runs. According to \eqref{eq:excessmse},
the excess MSE can be derived as $\xi_k = \text{tr} \left\{
\boldsymbol{R} E \big[ \boldsymbol{v}_{k} \boldsymbol{v}^T_k \big]
\right\}$. In this simulation example with $\boldsymbol{x}_k$
being an i.i.d. binary phase-shift keying (BPSK) sequence, the
covariance matrix $\boldsymbol{R}$ becomes identity, and
therefore, $\xi_k$ can be evaluated as $\text{tr} \left\{ E \big[
\boldsymbol{v}_{k} \boldsymbol{v}^T_k \big] \right\}$.

\begin{figure}%[t]
\centering
\includegraphics[scale = 0.48]{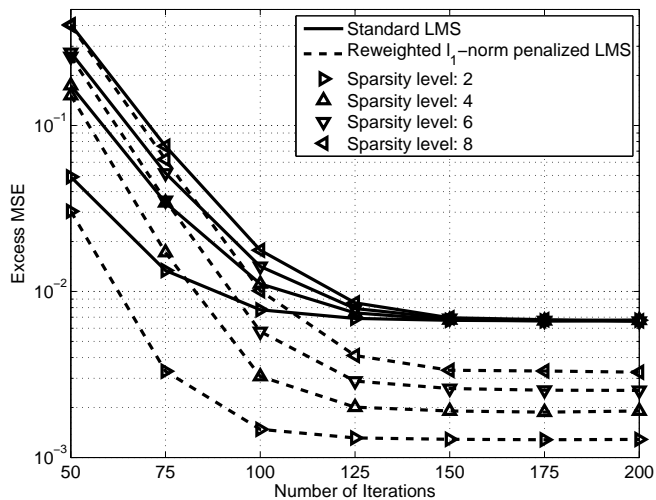}
\caption{Example 3: Excess MSE versus number of iterations.}
\label{fig:sim3}
\end{figure}

Fig.~\ref{fig:sim3} shows the excess MSE versus the number of
iterations for the standard LMS and reweighted $l_1$-norm
penalized LMS algorithms when the CIR sparsity level is varied
from $2$ to $8$. It can be seen that the standard LMS algorithm
results in the same excess MSE regardless of the sparsity level of
the CIR. However, the excess MSE of the reweighted $l_1$-norm
penalized LMS algorithm increases with increasing sparsity level
which is due to the fact that the value of $\alpha^{\prime}$ in
equation \eqref{eq:alphaPrime} is decreasing. For example,
$\alpha^{\prime}$ is equal to $2.7$, $2.3$, $2.0$, and $1.6$ for
sparsity levels of 2, 4, 6, and 8, respectively, after $150$
iterations. It can be also seen that in all cases, the reweighted
$l_1$-norm penalized LMS algorithm outperforms the standard LMS
algorithm.

\section{Conclusions}
\label{sec:Conc} Sparse channel estimation problem has been
considered in this paper and the reweighted $l_1$-norm penalized
LMS algorithm has been introduced and analyzed.
Quantitative analysis of the reweighted
$l_1$-norm penalized LMS algorithm and the
attainable excess MSE have been presented. The excess MSE result
shows that the reweighted $l_1$-norm penalized LMS algorithm
outperforms the standard LMS algorithm for the case of sparse CIR. 
The analysis has enabled us also to answer the
question of what is the maximum sparsity level in the channel for
which the reweighted $l_1$-norm penalized LMS algorithm is better
than the standard LMS. Update equations of the reweighted
$l_1$-norm penalized LMS, ZA-LMS, and the $l_p$-pseudo-norm
penalized LMS algorithms have been generalized to the case of an
arbitrary sparsity basis. Simulation results for the DCT sparse
channel are given along with simulation results for the time
sparse channel. The performance of the reweighted $l_1$-norm
penalized LMS algorithm has been compared to that of the standard
LMS, ZA-LMS, RZA-LMS algorithms, and our earlier proposed
$l_p$-pseudo-norm penalized LMS algorithm through computer
simulations. These results show that the reweighted $l_1$-norm
penalized LMS algorithm outperforms the standard LMS, ZA-LMS, and
RZA-LMS algorithms in all examples. It is also worth mentioning 
that variable step size is known to lead to better steady state 
error and therefore, better performance. Thus, as a further 
extention, the variable step size feature can be easily added 
to the proposed algorithm in the same way as it has been added 
to the RA-LMS in \cite{RewRef}. 

\footnotesize
%\biboptions{numbers,sort&compress}
%\bibliographystyle{elsarticle-num}

\end{document}